\documentclass[aps,prd,twocolumn,showpacs,superscriptaddress,groupedaddress,nofootinbib,floatfix]{revtex4-1}

\usepackage{amsmath}
\usepackage{amssymb}
\usepackage{graphicx}
\usepackage{hyperref}
\usepackage{color}
\usepackage{lineno}

\begin{document}

\title{Learning to Unscramble: \\
Simplifying Symbolic Expressions via Self-Supervised Oracle Trajectories}

\author{David Shih}
\email{shih@physics.rutgers.edu}
\affiliation{NHETC, Dept.\ of Physics and Astronomy, Rutgers University, Piscataway, NJ 08854, USA}

\date{\today}

\begin{abstract}
We present a new self-supervised machine learning approach for symbolic simplification of complex mathematical expressions. Training data is  generated  by scrambling simple expressions and recording the inverse operations, creating oracle trajectories that provide both goal states and explicit paths to reach them. A permutation-equivariant, transformer-based policy network is then trained on this data step-wise to predict the oracle action given the input expression. We demonstrate this approach on two problems in high-energy physics: dilogarithm reduction and spinor-helicity scattering amplitude simplification. In both cases, our trained policy network achieves near perfect solve rates across a wide range of difficulty levels, substantially outperforming prior approaches based on reinforcement learning and end-to-end regression. When combined with contrastive grouping and beam search, our model achieves a 100\% full simplification rate on a representative selection of 5-point gluon tree-level amplitudes in Yang-Mills theory, including expressions with over 200 initial terms.
\end{abstract}

\maketitle

\section{Introduction}

Machine learning (ML) methods have shown increasing promise across a range of symbolic mathematical tasks, including symbolic integration~\cite{Lample2019}, symbolic regression~\cite{Cranmer2020}, theorem proving~\cite{Lample2022}, symbolic equation solving~\cite{Dabelow2024}, and mathematical conjecture generation~\cite{Davies2021}. In this work, we focus on the task of \emph{symbolic simplification}: reducing complex mathematical expressions into compact, interpretable forms. Discovering the sequence of algebraic manipulations needed to simplify an expression is a fundamental challenge across countless fields, and no general-purpose algorithm exists. The difficulty is combinatorial: at each step, many possible identities could be applied to many parts of the expression, and the right choice often temporarily increases complexity before subsequent cancellations yield a simpler form.

Theoretical physics provides particularly compelling examples of this challenge, where simplification is often expected on the basis of symmetries and can be essential for extracting physical insight. In this paper, we study the application of machine learning (ML) techniques to two such examples: simplifying dilogarithm identities~\cite{Kirillov1995} that arise in one-loop and higher-order Feynman integrals~\cite{Kotikov1991,GehrmannRemiddi2000}, and simplifying on-shell scattering amplitudes~\cite{ParkeTaylor,Nair1988,Witten:2003nn,Cachazo:2004kj,Britto:2005fq} expressed in the spinor-helicity formalism~\cite{DixonTASI,ElvangHuang,CheungTASI}. These problems were first introduced as targets for ML methods by Dersy, Schwartz, and Zhang~\cite{Dersy:2022ltg} (hereafter DSZ) and Cheung, Dersy, and Schwartz~\cite{Cheung:2024tbb} (hereafter CDS), respectively. In both works, the strategy was to start from simplified expressions, apply a sequence of mathematical identities to ``scramble'' them into complicated expressions, and then train a seq2seq transformer to regress the simplified form from the scrambled input. DSZ also explored reinforcement learning (RL) for dilogarithm identities, which did not perform as well as seq2seq regression.

In this work, we revisit these problems and show how a new approach can significantly improve on the previous state-of-the-art, achieving nearly perfect simplification rates for both dilogarithms and spinor-helicity amplitudes. As in RL, our approach formulates the simplification task as a Markov decision process (MDP), in which the state is the current expression, the actions are identity applications, and transitions are deterministic. However, unlike RL, we do not train the MDP to explore the state space in order to maximize a reward function. Instead, we train it to exactly reproduce ``oracle trajectories" that transform complicated expressions to simple ones step-by-step. A transformer-based policy network is trained on these trajectories to predict, at each step, which identity to apply to which part of the expression.

The key insight of this work is that these oracle trajectories can be produced cheaply and efficiently, by first generating random scrambles of simple expressions into complicated ones (as in \cite{Dersy:2022ltg,Cheung:2024tbb}), {\it and then reversing them}. While simplification is hard (the search space of possible identity applications is combinatorially large), \emph{complexification} is easy: one can always make an expression more complicated by applying identities in the forward direction. By recording the inverse operations during forward scrambling, we can obtain a near limitless supply of oracle trajectories. 

Since the training data is generated without any expert knowledge, our approach is an example of {\it self-supervised} training. This distinguishes our approach from that of behavioral cloning~\cite{Pomerleau1988,Bain1995}, where demonstrations are generated by a human expert. The same self-supervised idea was employed by Agostinelli et al.~\cite{Agostinelli2019} and Takano et al.~\cite{RubiksCube} for solving Rubik's cube; a notable difference is that our problems can have multiple equivalent goal states rather than one canonical solved state.

Several technical innovations prove essential for high performance. We introduce a \emph{multi-label soft loss} to handle action equivalence: algebraic symmetries of the underlying identities often mean that multiple distinct actions produce the same simplified result, and penalizing the model for choosing a valid alternative would be counterproductive. The transformer policy network is fully permutation equivariant with respect to the terms in the input expressions, reflecting the fact that terms in a mathematical expression form an unordered set. At inference time, anti-cycle detection, backtracking, and term-count constraints further boost performance.

We demonstrate our approach on both problems described above, benchmarking our performance on the same test sets used by DSZ and CDS respectively to enable head-to-head comparisons.  For dilogarithm simplification, we achieve a \textbf{99.9\%} solve rate (4,731/4,737 on the test set of DSZ, after excluding overlap and pathological samples), compared to 92\% for DSZ's best method. For scattering amplitudes, we achieve \textbf{99.9\%} on 4-point, \textbf{99.6\%} on 5-point, and \textbf{99.4\%} on 6-point amplitudes under a target-relative criterion (output as simple as the known simplified form), compared to 98.2\% (4-point), 96.0\% (5-point), and 96.9\% (6-point) for the model of CDS under the same criterion. These amount to reductions of the failure rate ranging from factors of 5 to 80 across the board.

Finally, we go beyond the test sets of CDS (which average ${\sim}$4--10 source terms) and tackle a much harder challenge (also explored in CDS): simplifying actual tree-level 5-point gluon amplitudes in Yang-Mills theory. Depending on the choice of external polarizations, when calculated with Feynman diagrams, these range from 8 to 228 terms (mean ${\sim}$90) and are all equivalent to the single-term Parke-Taylor formula \cite{ParkeTaylor}. Since these expressions far exceed the model's 25-term input capacity, we combine our trained MDP with contrastive grouping~\cite{Cheung:2024tbb} to decompose large expressions into manageable sub-problems, and embed the result in a beam search to navigate the vast combinatorial space of identity sequences. This pipeline achieves a \textbf{100\%} solve rate on a representative subset of 103 forms, demonstrating that the MDP trained on simple scrambles can generalize to simplify realistic Feynman-diagram-level amplitudes when augmented with appropriate search strategies.

The remainder of this paper is organized as follows. Section~\ref{sec:framework} describes the general framework: MDP formulation, oracle trajectory generation, multi-label loss, architecture, and inference techniques. Sections~\ref{sec:dilog} and~\ref{sec:amp} apply this framework to dilogarithms and scattering amplitudes, respectively, presenting problem-specific details and results. Section~\ref{sec:discussion} discusses and concludes.

\section{Method}
\label{sec:framework}

We now describe the general framework underlying our approach. The key components---MDP formulation, oracle trajectory generation, multi-label loss, architecture, and inference techniques---are presented here in problem-agnostic terms. Sections~\ref{sec:dilog} and~\ref{sec:amp} illustrate this framework for the two specific problems studied in this work.

\subsection{Symbolic Simplification as an MDP}

We consider simplification problems with the following structure. The \emph{state} is a symbolic mathematical expression (e.g., a sum of special functions or a rational function of spinor brackets). The \emph{actions} are applications of mathematical identities to specific parts of the expression (e.g., applying a functional identity to a particular term). The dynamics are deterministic: applying an identity to an expression produces a unique new expression. The \emph{goal} is to reach a state where the expression is ``simple'' according to a problem-specific criterion (e.g., minimal number of terms or brackets).

This naturally defines a Markov decision process (MDP) $(S, A, T, G)$ where $S$ is the set of all valid expressions, $A$ is the set of identity applications, $T: S \times A \to S$ is the deterministic transition function, and $G \subset S$ is the set of goal states. The policy $\pi(a|s)$ maps states to distributions over actions; we seek a policy that, when applied greedily step by step, navigates from any given expression to a simplified form. 

Two features distinguish this MDP from typical settings. First, the goal set $G$ typically contains many equivalent simplified forms rather than a single canonical target---any expression meeting the simplicity criterion is acceptable. Second, algebraic symmetries often mean that multiple distinct actions from a given state produce the same result (Section~\ref{sec:multilabel}). Both features require careful treatment in the loss function and evaluation.

\subsection{Oracle Trajectory Generation}
\label{sec:oracle_gen}

While \emph{simplification} is hard (the search space is combinatorially large), \emph{complexification} is easy: one can always make an expression more complicated by applying identities in the forward direction. The central insight of our approach is that this asymmetry enables automatic generation of oracle trajectories.

The procedure has three steps:
\begin{enumerate}
\item \textbf{Construct a simple target expression} from the goal set $G$, following problem-specific generation rules (see Sections~\ref{sec:dilog_data} and~\ref{sec:amp_data}).

\item \textbf{Scramble forward.} Apply a sequence of $n_{\rm scr}$ random identity transformations to produce a complex expression. At each step, select a random part of the expression and apply a randomly chosen identity. Record the sequence of intermediate states: $[s_0, s_1, \ldots, s_{n_{\rm scr}}]$, where $s_0$ is the simple target and $s_{n_{\rm scr}}$ is the fully scrambled expression.

\item \textbf{Reverse to obtain oracle trajectory.} Reverse the state sequence to get $[s_{n_{\rm scr}}, \ldots, s_1, s_0]$---a trajectory from complex to simple. For each consecutive pair $(s_i, s_{i-1})$, find the inverse action by brute-force search: try all valid identity applications on $s_i$ and identify which one(s) produce $s_{i-1}$. This yields oracle tuples $(s_t, a_t, s_{t+1})$ providing explicit step-by-step guidance.
\end{enumerate}

This procedure is fully automatic and can generate unlimited training data. As noted in the Introduction, it is closely related to behavioral cloning~\cite{Pomerleau1988,Bain1995}, with the crucial distinction that demonstrations are generated synthetically rather than by a human expert. 
A well-known limitation of behavioral cloning is compounding errors~\cite{Ross2011}: small mistakes push the agent into states not covered by the training data, causing cascading failures. In our setting, this is mitigated by two factors. First, the backward-scrambling procedure naturally generates training states at all difficulty levels, from nearly simplified to highly scrambled, providing dense coverage of the relevant state space. Second, the inference-time techniques described in Section~\ref{sec:inference_techniques} (anti-cycle detection, backtracking) enable recovery from mistakes.

\subsection{Multi-Label Soft Loss}
\label{sec:multilabel}

In many symbolic domains, algebraic symmetries mean that multiple distinct actions can produce the same result from a given state. If we train with standard single-label cross-entropy, choosing any one oracle action as the ``correct'' label penalizes the model for predicting equally valid alternatives.

To handle this, we use a multi-label soft loss. Given a state $s$ with equivalent oracle actions $\{a_1, \ldots, a_k\}$, each receives target probability $1/k$:
\begin{equation}
\mathcal{L} = -\sum_{i=1}^{k} \frac{1}{k} \log \pi(a_i | s)
\label{eq:multilabel}
\end{equation}
This distributes credit equally across all valid actions, teaching the model to recognize multiple valid simplification strategies. When actions are unique ($k=1$), this reduces to standard cross-entropy.

The impact of multi-label loss can be dramatic. For scattering amplitude simplification, where the Schouten identity creates on average ${\sim}6.2$ equivalent inverse actions per transition, switching from single-label to multi-label loss improves the 4-point solve rate from 73\% to 98\%. For problems where action equivalence does not arise (e.g., dilogarithm simplification, where each identity--term pair produces a unique result), standard single-label cross-entropy suffices.

\subsection{Architecture}
\label{sec:architecture}

\begin{figure}[t]
\centering
\includegraphics[width=\columnwidth]{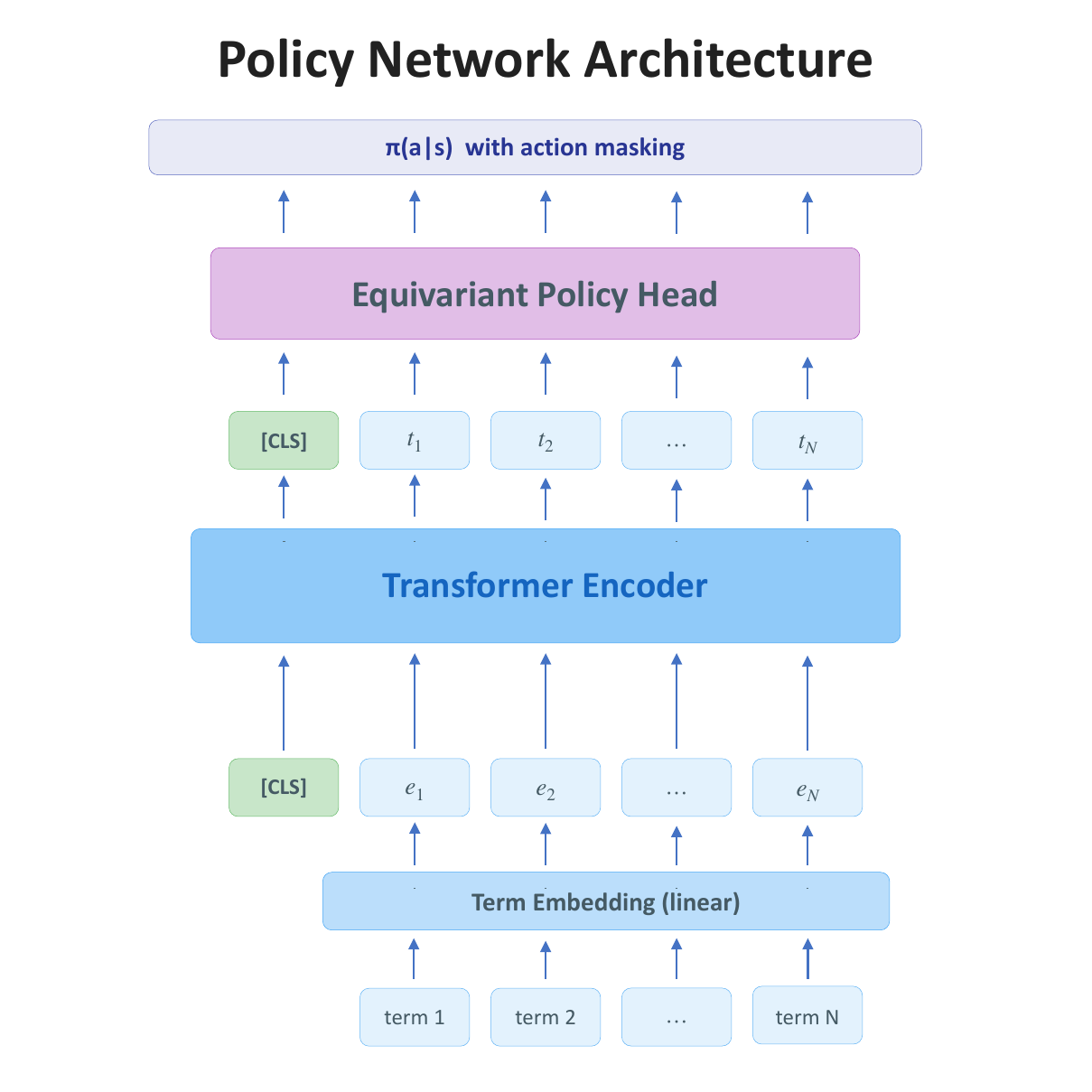}
\caption{Architecture of the policy network for symbolic simplification. Each term's feature vector is embedded and processed by a Transformer encoder with a prepended learnable \text{[CLS]} token. No positional encoding is used, respecting the permutation symmetry of terms. The permutation equivariant policy head takes the output of the transformer and returns probabilities (softmax) over the action space.}
\label{fig:architecture}
\end{figure}

For the policy network $\pi(a|s)$, we use a transformer encoder~\cite{Vaswani2017} that takes a state $s$ as input and outputs a softmax over all the actions $a$ in the action space. Our transformer encoder has the following design choices motivated by the structure of the simplification problem (Fig.~\ref{fig:architecture}):

\textbf{Input representation.} Each term in the expression is encoded as a fixed-dimensional feature vector using a problem-specific encoding (see Sections~\ref{sec:dilog} and~\ref{sec:amp}). The expression is represented as a padded sequence of up to $N_{\max}$ such term vectors.

\textbf{No positional encoding.} The terms in a mathematical expression form an unordered set (addition is commutative), so positional encoding would impose a spurious ordering. We omit it entirely.

\textbf{Global summary token.} A learnable \texttt{[CLS]} token is optionally prepended to the term sequence. Its output embedding aggregates global information about the expression via self-attention and serves as a summary representation that is permutation invariant.

\textbf{Policy head.} The structure of the policy head depends on the action space. When actions decompose as (identity, term)---i.e., which identity to apply to which term---we use an \emph{equivariant} per-term policy head: a shared MLP is applied independently to each term's contextualized output embedding to produce logits for the identities applicable to that term. This ensures equivariance under permutations of the input terms. When actions have a different structure (e.g., specified by bracket indices rather than terms), a global policy head operating on the aggregated CLS representation is used instead. Problem-specific details are given in Sections~\ref{sec:dilog} and~\ref{sec:amp}.

\textbf{Action masking.} Actions targeting non-existent terms (padding) or invalid operations are masked by setting their logits to $-\infty$ before the softmax, ensuring zero probability for invalid actions.

The specific architecture used for all problems in this work has embedding dimension 64, 4 attention heads, 3 layers, and feed-forward dimension 128, totaling ${\sim}165$K parameters. Training uses AdamW (learning rate $10^{-4}$, linear warmup over 5 epochs, weight decay 0.01, gradient clipping at max norm 1.0) for 100--400 epochs.

\subsection{Inference Techniques}
\label{sec:inference_techniques}

At inference time, the trained policy is applied greedily step by step. Several techniques improve robustness beyond naive greedy rollout:

\textbf{Anti-cycle detection.} We maintain a set of visited (state-hash, action) pairs throughout each episode. Any action previously taken from the current state is masked, forcing exploration of alternative paths. 
This prevents the policy from entering infinite loops (e.g., applying an identity and then its inverse repeatedly).

\textbf{Backtracking.} We checkpoint states at locally minimal complexity (e.g., minimal bracket count). If the episode reaches the maximum step limit or times out without solving, execution returns to the best checkpoint and tries the next-best action according to the policy. Up to $M$ backtrack attempts are allowed (typically $M{=}10$), each exploring a different action from the checkpoint state. For amplitude simplification, backtracking recovers ${\sim}$72\% of initial failures (Section~\ref{sec:amp_eval}).

\textbf{Reject Term Increase (RTI).} Actions that would increase the term count by more than a threshold $N$ are masked. This prevents the model from exploring branches that lead to term explosion, which becomes increasingly problematic at higher multiplicities where the combinatorial space of identity applications is larger. The optimal threshold is problem-dependent: no RTI may be needed when the action space is small, while tight constraints (e.g., $N{=}1$) may be necessary for large action spaces where even moderate term increases tend to be irrecoverable.

Not all techniques are required for every problem. For dilogarithm simplification, anti-cycle detection alone achieves 99.9\% solve rate. For scattering amplitudes, the full suite is used, with problem-specific settings detailed in Section~\ref{sec:amp}.

\subsection{Success Criteria}
\label{sec:success_criteria}

When evaluating simplification models, it is important to distinguish two natural success criteria that apply across problems:
\begin{itemize}
\item \textbf{Source-relative:} The output has strictly lower complexity than the input (source). This verifies that the model has achieved at least some simplification. The complexity measure is problem-specific: for dilogarithms, it is the number of terms; for amplitudes, the output must be strictly simpler in at least one dimension---i.e., (bracket count $<$ source \emph{and} term count $\leq$ source) or (bracket count $\leq$ source \emph{and} term count $<$ source).
\item \textbf{Target-relative:} The output has complexity at or below that of the known simplified form (target). This is a stricter criterion that verifies the model has reached the level of simplification achieved by the known answer. For dilogarithms, this means the term count reaches the number of terms in the target or fewer; for amplitudes, both the bracket count and the term count must be $\leq$ the target's values.
\end{itemize}
Source-relative success is a weaker condition: the model may simplify somewhat without reaching the full simplification achievable. Target-relative success is our primary metric, as it measures whether the model truly solves the simplification problem.

\section{Dilogarithm Simplification}
\label{sec:dilog}

\subsection{Background}

The dilogarithm function is defined by
\begin{equation}
\text{Li}_2(x) = -\int_0^x \frac{\ln(1-t)}{t}\, dt = \sum_{n=1}^{\infty} \frac{x^n}{n^2}.
\end{equation}
Dilogarithms appear ubiquitously in loop-level calculations in quantum field theory, particularly in Feynman integrals~\cite{Kotikov1991,GehrmannRemiddi2000}. Expressions involving sums of dilogarithms often simplify dramatically through functional identities~\cite{Zagier2007,Kirillov1995}. Working modulo non-dilogarithmic terms (constants and products of logarithms), the relevant identities are:
\begin{align}
\text{Li}_2(z) &= -\text{Li}_2(1-z) + \cdots &\text{(reflection)} \label{eq:reflection} \\
\text{Li}_2(z) &= -\text{Li}_2(1/z) + \cdots &\text{(inversion)} \label{eq:inversion} \\
\text{Li}_2(z) &= -\text{Li}_2(-z) + \tfrac{1}{2}\text{Li}_2(z^2) &\text{(duplication)} \label{eq:duplication}
\end{align}
where the ellipsis denotes terms involving $\pi^2$, $\ln$, and constants that we drop since we track only the dilogarithmic content. Note that reflection and inversion preserve the number of terms, while duplication replaces one term with two.

We consider expressions of the form
\begin{equation}
E(x) = \sum_{i=1}^{N} c_i\, \text{Li}_2\!\big(h_i(x)\big),
\label{eq:dilog_expr}
\end{equation}
where the $c_i$ are rational coefficients and each $h_i(x)=p_i(x)/q_i(x)$ is a rational function of a single variable $x$ with integer polynomial coefficients. The simplification goal is to reduce $E(x)$ to the fewest possible dilogarithm terms---ideally zero if the expression is a sum that cancels exactly.

DSZ studied this problem using both reinforcement learning (TRPO) and sequence-to-sequence transformers. Their RL agent trained on ``easy" expressions that simplify to zero achieved 50--59\% greedy solve rate and up to 89\% with beam search on a held out test set. Notably, both of these solve rates were strictly {\it worse} than a simple classical greedy best-first search algorithm that achieved 91\% on the same test set. On a more difficult setting that included nonzero expressions and more scrambles, they reported a 92\% solve rate using an end-to-end transformer (after canonicalizing equivalent forms).

\subsection{Data Generation}
\label{sec:dilog_data}

Training data is generated by constructing expressions that are \emph{known} to simplify, then recording the inverse operations as oracle trajectories. The procedure is:

\begin{enumerate}
\item \textbf{Construct a simple target expression.} We follow the data generation procedure of DSZ, using the same parameter ranges. Each initial argument $h_i(x) = p_i(x)/q_i(x)$ has $p_i, q_i$ with integer coefficients in $[-2,2]$ and degrees in $\{0,1,2\}$. Expressions are built from two ingredients: $n_s$ ``base'' dilogarithm terms $c_j\,\text{Li}_2(h_j(x))$ with $c_j \in [-8,-1] \cup [1,8]$ that form the simplified target, plus $n_t$ ``zero pairs'' $c_i\,\text{Li}_2(h_i(x)) - c_i\,\text{Li}_2(h_i(x)) = 0$ with $c_i \in [1, 8]$ that vanish upon simplification. When $n_s = 0$, the target is zero.

\item \textbf{Scramble forward.} Apply $n_{\text{scr}} \in [1, 7]$ random identity transformations. At each step, select a random dilogarithm term in the expression and apply a randomly chosen identity (reflection, inversion, or duplication). To prevent trivial undoing, the same identity cannot be applied consecutively to a term with the same argument. After each identity application, terms with matching arguments are combined (coefficients summed) and zero-coefficient terms are removed.

\item \textbf{Reverse and find inverse actions} following the procedure of Section~\ref{sec:oracle_gen}. For each consecutive pair $(s_i, s_{i-1})$, the inverse action is found by trying all $3 \times N_{\text{terms}}$ possible (identity, term) combinations on $s_i$. Action equivalence does not arise here, as each identity--term pair produces a unique result.
\end{enumerate}

We generate 100k training samples, split equally across four target complexities $n_s \in \{0, 1, 2, 3\}$ (25k samples each), where $n_s$ is the number of terms in the simplified target expression. Each configuration uses $n_t$ zero pairs and a maximum of $n_{\text{scr}}$ scrambles: ($n_s=0$: $n_t=2$, $n_{\text{scr}} \leq 5$), ($n_s=1$: $n_t=1$, $n_{\text{scr}} \leq 5$), ($n_s=2$: $n_t=1$, $n_{\text{scr}} \leq 6$), ($n_s=3$: $n_t=1$, $n_{\text{scr}} \leq 7$). 

\subsection{MDP Formulation and Architecture}

\textbf{State representation.} Each dilogarithm term $c_i\,\text{Li}_2(h_i(x))$ is encoded as a 17-dimensional feature vector: $\log(1+|c_i|)$, $\text{sign}(c_i)$, the numerator degree, denominator degree, up to 6 numerator polynomial coefficients (log-transformed; truncated from the lowest degree if the polynomial exceeds degree 5), up to 6 denominator polynomial coefficients (same encoding), and a padding flag indicating whether the term slot is occupied (1) or empty (0). The expression is represented as an ordered sequence of up to 15 such term vectors. 

\textbf{Action space.} There are $3 \times 15 = 45$ discrete actions, corresponding to applying one of 3 identities (reflection, inversion, duplication) to one of up to 15 terms. Actions targeting non-existent terms are masked (logits set to $-\infty$).

\textbf{Architecture.} The policy network is a Transformer encoder (embedding dimension 64, 4 attention heads, 3 layers, feed-forward dimension 128, ${\sim}$165K parameters), illustrated in Fig.~\ref{fig:architecture}. No positional encoding is used, since the terms are an unordered set. The policy head is \emph{equivariant}: an MLP is applied independently to each term's output embedding to produce 3 logits (one per identity), yielding a $15 \times 3 = 45$-dimensional action distribution. This ensures that the logit for applying identity $a$ to term $i$ depends on term $i$'s contextualized representation, not on its position in the sequence.

\textbf{Training.} Standard cross-entropy loss on oracle actions, using the optimizer settings described in Section~\ref{sec:architecture}. We train for 200 epochs on the 100k oracle trajectories, which expand to ${\sim}$478k individual state-action transitions. Action equivalence does not arise for dilogarithms (each identity--term pair produces a unique result), so single-label cross-entropy suffices.

\textbf{Inference.} Greedy action selection with anti-cycle detection: we maintain a set of visited (state-hash, action) pairs and mask any action previously taken from the current state, forcing exploration of alternative paths. An episode succeeds if the number of terms reaches the target value ($\leq n_s$) at any point during the rollout.

\subsection{Results}

We evaluate our trained model on the test set of DSZ, which contains 5,000 expressions with $n_s \in \{0,1,2,3\}$ target terms, scrambled up to 10 times. We exclude 164 test expressions (3.3\%) that have exact matches in our training data---including all intermediate states visited during oracle trajectory generation, not just the starting expressions---accounting for term reordering (which is a symmetry of our permutation-equivariant model), and 128 pathological expressions where the source has no more terms than the target (making simplification undefined). After removing the overlap between these two sets (29 samples), we are left with 4,737 clean expressions. 
For comparison, we also re-evaluated the pretrained seq2seq transformer of DSZ~\cite{DSZcode} (greedy decoding) on the same clean test set.

Table~\ref{tab:dilog_results} shows results under both the source-relative and target-relative criteria defined in Section~\ref{sec:success_criteria}, broken down by target complexity and scramble depth.  For dilogarithms, source-relative success means any reduction in term count from the starting expression, while target-relative success means the term count reaches the number of terms in the initial unscrambled expression or fewer.  For the DSZ model we only report one set of results, as there is no difference between the source-relative and the target-relative criteria -- either the model produces the target expression exactly, or it hallucinates a mathematically inequivalent expression. It never produces a partially-simplified but mathematically-equivalent expression.

\begin{figure}[t]
\centering
\includegraphics[width=\columnwidth]{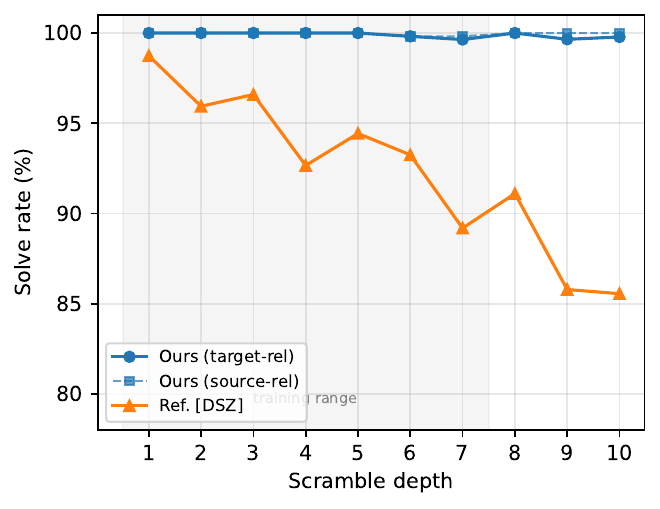}
\caption{Solve rate vs.\ scramble depth for dilogarithm simplification. Our model (blue) maintains near-100\% performance under both source-relative and target-relative criteria, even beyond the training range (shaded). The seq2seq model of DSZ (orange) degrades at higher scramble depths.}
\label{fig:dilog_solve_rate}
\end{figure}

Under the looser source-relative criterion, our model achieves a failure rate of only $4.2\times 10^{-4}$ (2 failures out of 4,737 total examples). Under the tighter target-relative criterion, this goes up to $1.3\times 10^{-3}$ (6 failures) -- the 4 additional target-relative failures achieve partial simplification. By contrast, we find the seq2seq model of DSZ fails to simplify 8.4\% of the expressions (399 out of 4,737), a much higher failure rate than our approach. The breakdown in Table~\ref{tab:dilog_results} reveals where their model struggles: performance drops sharply with increasing target complexity, from 99.9\% at $n_s{=}0$ to 86.4\% at $n_s{=}3$, and with increasing scramble depth, from ${\sim}$96\% at depths 1--5 to ${\sim}$86\% at depths 9--10. In contrast, our model maintains $\geq$99.6\% across all categories under both criteria. This is further illustrated in Figure~\ref{fig:dilog_solve_rate}, which shows the solve rate as a function of scramble depth for both models. It is highly notable that our model was only trained on $\le 7$ scrambles, yet continues with no loss in performance out to 10 scrambles. This is a strong sign that our model is truly learning an effective stepwise reduction strategy which it can then generalize successfully beyond the limitations of the training set. 

Further analysis of the 6 target-relative failures reveals that they span all target complexities and occur at scramble depths $\geq$6. The complete source, target, and model output expressions for each failure are listed in Appendix~\ref{app:dilog_failures}. Failure 3 is notable: the source has only 4 terms but with arguments involving degree-8 polynomials, and the model's attempts to simplify actually {\it increase} the term count (ending at 8 terms after 50 steps). Failure 6 is striking: the target is exactly zero (a pure identity), yet our model reduces 5 terms to 2 without fully canceling them. Per-sample comparison with DSZ reveals that our model solves 398 of their 399 failures on the clean set, while they solve 5 of our 6; only 1 expression defeats both models.

\begin{figure}[t]
\centering
\includegraphics[width=\columnwidth]{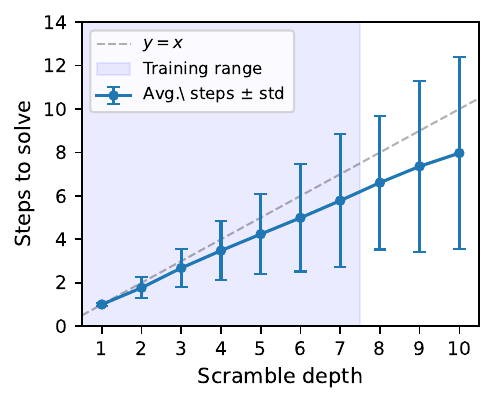}
\caption{Average number of steps to solve vs.\ scramble depth for dilogarithm simplification, with $\pm$1 standard deviation error bars. The dashed line marks $y=x$ (steps equal to scramble depth); the shaded region indicates the training range (scramble depths 1--7). The model consistently finds shorter paths than the scramble depth, indicating it learns to bypass redundancy introduced by the scrambling process.}
\label{fig:avg_steps}
\end{figure}

Finally, in Fig.~\ref{fig:avg_steps} we show statistics on the number of steps our model takes to fully simplify a source expression vs.\ the number of scrambles that produce the source expression in the first place.  Remarkably, the model finds {\it shorter} simplification paths than the number of scrambles on average: the ratio of average steps to scramble depth decreases from 0.90 at depth 2 to 0.80 at depth 10, indicating that the scrambling process introduces redundancy that the model learns to bypass.

\begin{table}[t]
\centering
\begin{tabular}{lc cc c}
\hline\hline
& & \multicolumn{2}{c}{Our model} & DSZ \\
& $N$ & Source-rel & Target-rel &  \\
\hline
\multicolumn{5}{l}{\emph{By target terms}} \\
 0  & 810   & 0 & 1 & 1 \\
 1  & 902   & 1 & 2 & 41 \\
 2  & 1,160 & 1 & 2 & 104 \\
 3  & 1,865 & 0 & 1 & 253 \\
\hline
\multicolumn{5}{l}{\emph{By scramble depth}} \\
 1--5 & 2,043 & 0 & 0 & 104 \\
 6    & 546   & 1 & 1 & 37 \\
 7    & 554   & 1 & 2 & 60 \\
 8    & 562   & 0 & 0 & 50 \\
 9    & 577   & 0 & 2 & 82 \\
 10   & 455   & 0 & 1 & 66 \\
\hline
Total    & 4,737 & 2 & 6   & 399   \\
\hline\hline
\end{tabular}
\caption{Dilogarithm simplification failures on the clean test set of DSZ (4,737 samples after excluding 164 overlapping with our training data and 128 pathological, with 29 in both), broken down by number of initial terms and by the number of scrambles. Source-relative: failure to produce a correct output with strictly fewer terms than the initial scrambled expression. Target-relative: failure to reach term count of the initial unscrambled expression. For the seq2seq model of DSZ, we only report one column of failure numbers as there is no empirical difference between the two criteria: it turns out that their model either outputs expressions that exactly reproduce the target, or it outputs incorrect outputs that are not mathematically equivalent to the input.}
\label{tab:dilog_results}
\end{table}

\section{Scattering Amplitude Simplification}
\label{sec:amp}

\subsection{Background}

Scattering amplitudes in massless gauge theories are most compactly expressed using the spinor-helicity formalism~\cite{DixonTASI,ElvangHuang,CheungTASI}. External momenta are encoded via spinors, with two fundamental products:
\begin{align}
\langle ij \rangle &= \bar{u}_-(p_i) u_+(p_j) \quad \text{(angle bracket)} \\
[ij] &= \bar{u}_+(p_i) u_-(p_j) \quad \text{(square bracket)}
\end{align}
While Feynman diagram calculations  can easily yield hundreds of terms, these are often found to simplify drastically. The most celebrated example is the Parke-Taylor formula~\cite{ParkeTaylor} for $n$-point MHV gluon scattering:
\begin{equation}
A(1^+\cdots i^-\cdots j^-\cdots n^+) = \frac{\langle ij \rangle^4}{\langle 12 \rangle \langle 23 \rangle \cdots \langle n1 \rangle}
\label{eq:parke_taylor}
\end{equation}
More generally, on-shell recursion relations~\cite{Britto:2005fq} prove that all tree-level amplitudes can be constructed from on-shell building blocks without reference to Feynman diagrams, guaranteeing the existence of compact representations. This makes tree-level amplitudes a good testing ground for methods that systematically reduce complicated Feynman-diagram-level expressions to their simplest equivalent forms. 

At tree-level, simplification relies on three families of identities:
\begin{align}
\langle ij \rangle \langle kl \rangle &= \langle il \rangle \langle kj \rangle + \langle ik \rangle \langle jl \rangle \quad &\text{(Schouten)} \label{eq:schouten} \\
\textstyle\sum_{j} \langle ij \rangle [jk] &= 0 \quad &\text{(mom.\ cons.)} \label{eq:momcons} \\
\textstyle\sum_{i<j \in S_1} \langle ij \rangle [ji] &= \textstyle\sum_{k<l \in S_2} \langle kl \rangle [lk] \quad &\text{(mom.\ squared)} \label{eq:momsq}
\end{align}
(and analogous identities with angle and square brackets exchanged). The Schouten identity applies to any pair of brackets sharing the same type, while momentum conservation and momentum squared involve sums over external particle indices. 

CDS studied this problem by using these identities to create simple/complex expression pairs through random scrambling operations, and then training
encoder-decoder transformers (seq2seq) to directly predict simplified outputs from complex inputs. They trained on 4-, 5-, and 6-point amplitudes with up to 3 scrambles and used beam search at inference to improve accuracy. As we discuss in Section~\ref{sec:amp_eval},  our model substantially outperforms theirs under both target-relative and source-relative criteria.

\subsection{Data Generation}
\label{sec:amp_data}

We use the data generation procedure of CDS, adapted to our oracle trajectory framework. As in the dilogarithm case, the key idea is to construct simple target expressions, scramble them forward using the identities, and reverse the trajectories to obtain oracle demonstrations. The procedure is:

\begin{enumerate}
\item \textbf{Construct a simple target expression.} Following CDS, each target is a rational function of spinor brackets with 1--3 additive terms sharing a common denominator. The first term is built by sampling $n_{\rm num}$ numerator and $n_{\rm den}$ denominator brackets (each drawn uniformly from $\{1,\ldots,2n\}$ where $n$ is the number of external particles), with each bracket randomly chosen as angle $\langle ij\rangle$ or square $[ij]$ (equal probability, $i\neq j$). Each bracket is raised to power $\max(1, \operatorname{Poisson}(\lambda{=}0.75))$, and the overall sign is $\pm 1$ with equal probability. For expressions with 2--3 terms, additional numerators are constructed by solving a Diophantine system to match the little-group scaling of the first term, ensuring physical consistency.

\item \textbf{Scramble forward.} Apply 1--3 random identity transformations (Schouten, momentum conservation, momentum squared). To increase training data diversity, we also include auxiliary scrambling operations: identity multiplication (multiply a term by $\langle ij\rangle[ji]/\langle ij\rangle[ji] = 1$, inserting new brackets without changing the value) and zero addition (add a term and its negation). These auxiliary operations produce a broader range of intermediate states that better cover the test distribution.

\item \textbf{Reverse and find inverse actions.} The scrambled trajectory is reversed, and inverse actions are found by brute-force search over all valid (identity, bracket, target) combinations. A key difference from the dilogarithm case is \emph{action equivalence}: each scrambling operation typically has multiple valid inverse actions due to algebraic symmetries. For example, applying the Schouten identity creates four cross brackets, and the reverse Schouten can be applied using any of them, yielding on average ${\sim}$6.2 equivalent inverse actions per transition. All equivalent actions are recorded for use with the multi-label loss.
\end{enumerate}

We generate 500k training trajectories per $n$-point problem; after expanding each multi-step trajectory into individual state-action transitions, this yields 950k (4-point), 944k (5-point), and 921k (6-point) training samples. For evaluation, we use the test sets published by CDS, which contain 10,020 (4-point), 10,270 (5-point), and 9,934 (6-point) samples. We checked for overlap between the scrambled test expressions and all states visited in our training trajectories, finding 28 matches for 4-point and 9 for 5-point. We also identified pathological samples where the target expression is not strictly simpler than the source under the source-relative criterion defined in Section~\ref{sec:amp_eval} (123 for 4-point, 5 for 5-point, 0 for 6-point). Both categories are excluded from evaluation.

\subsection{MDP Formulation and Architecture}

\textbf{State representation.} Each term in the amplitude is encoded as a feature vector. The first two components encode the numerical coefficient (log magnitude and sign). The remaining components describe up to $B$ brackets per term (padded or truncated): for each bracket, we store its type (angle or square), its power (log-scaled), and one-hot encodings of the two momentum arguments, giving $2 + 2n$ features per bracket where $n$ is the number of external particles. A validity flag indicates real vs.\ padded terms. The total per-term feature dimension is $2 + B(2+2n) + 1$; for 4-point amplitudes with $B{=}8$, this is 83. As in the dilogarithm case, the expression is represented as a padded sequence of up to $N_{\max}$ term vectors. We use $N_{\max} = 10$ for 4- and 5-point and $N_{\max} = 30$ for 6-point amplitudes.

\textbf{Action space.} Each action is specified by three components: (i) a \emph{bracket index} identifying which spinor bracket to target (from the canonical set of $\binom{n}{2} \times 2$ possible brackets), (ii) a \emph{term slot} specifying whether to apply the identity to the whole expression (slot 0) or to a specific term (slots $1, \ldots, N_{\max}$), and (iii) an \emph{identity type index} specifying which identity variant to use. For a bracket $\langle ij\rangle$ (or $[ij]$) with $n_{\rm other} = n - 2$ remaining momenta, there are $\binom{n_{\rm other}}{2}$ Schouten variants (one per pair of reference momenta), $2(1+n_{\rm other})$ momentum conservation variants (choosing which bracket argument to extract, and which momentum to use as denominator), and $2^{n_{\rm other}}$ momentum-squared variants (one per subset of remaining momenta). The total action space size is
\begin{equation}
|A| = n_{\text{brackets}} \times (1 + N_{\max}) \times n_{\text{id}},
\end{equation}
where $n_{\text{id}}$ is the number of identity types per bracket. The action space is fixed (independent of the current expression); invalid actions are masked at runtime. Concretely:
\begin{itemize}
\item 4-point: $12 \times 11 \times 11 = 1{,}452$ actions
\item 5-point: $20 \times 11 \times 19 = 4{,}180$ actions
\item 6-point: $30 \times 31 \times 32 = 29{,}760$ actions
\end{itemize}
The rapid growth is driven by all three factors: more brackets ($\sim n^2$), more term slots (larger expressions at higher multiplicities), and more identity parameters (more momenta available for momentum conservation and squared-momentum identities).

\textbf{Architecture.} The same Transformer encoder as for dilogarithms (embedding dimension 64, 4 attention heads, 3 layers, feed-forward dimension 128, ${\sim}$165K parameters), adapted for amplitude-specific term features. Unlike the dilogarithm case where the policy head is per-term equivariant (3 identity logits per term), the amplitude policy head operates on the global CLS token output and produces logits over the full action space of identity--bracket--term-slot combinations. This is because brackets span across terms, making the action space inherently global rather than per-term.

\textbf{Training.} The multi-label soft loss is essential here. For state $s$ with equivalent oracle actions $\{a_1, \ldots, a_k\}$, each receives target probability $1/k$:
\begin{equation}
\mathcal{L} = -\sum_{i=1}^{k} \frac{1}{k} \log \pi(a_i | s)
\end{equation}
Training uses the same optimizer settings as described in Section~\ref{sec:architecture}, for 200 epochs on 500k samples.

\textbf{Inference.} Greedy action selection with anti-cycle detection, as in the dilogarithm case. Two additional inference techniques are used for amplitudes:
\begin{itemize}
\item \textbf{Backtracking:} We checkpoint states at locally minimal complexity (bracket count). If the episode reaches the maximum step limit or times out without solving, we backtrack to the best checkpoint and try the next-best action. Up to 10 backtrack attempts are allowed. Only 0.8\% (4pt), 1.5\% (5pt), and 1.9\% (6pt) of samples require backtracking; when triggered, it recovers 84\%, 41\%, and 53\% of initial greedy failures to proper simplification, improving proper solve rates by 0.7, 0.6, and 1.0 percentage points respectively.
A per-step timeout of 30 seconds and a per-sample timeout of 120--180 seconds (depending on $n$-point) protect against expensive identity applications.
\item \textbf{Reject Term Increase (RTI):} Actions that would increase the term count by more than a threshold $N$ are masked. This prevents the model from exploring branches that lead to term explosion, which becomes increasingly problematic at higher multiplicities where the combinatorial space of identity applications is larger.  We have found empirically that $N{=}1$ is optimal across all multiplicities, with the most dramatic effect at 6-point. We adopt RTI with $N=1$ for all subsequent evaluations of our model.

\end{itemize}

\subsection{Results}
\label{sec:amp_eval}

We evaluate our trained model on the test sets published by CDS, which contain 10,020 (4-point), 10,270 (5-point), and 9,934 (6-point) samples. We exclude pathological samples where the target is not strictly simpler than the source under the source-relative criterion (123 for 4-point, 5 for 5-point, 0 for 6-point), as well as the train/test overlap samples identified in Section~\ref{sec:amp_data} (28 for 4-point, 9 for 5-point, 0 for 6-point), leaving clean test sets of 9,869, 10,256, and 9,934 expressions respectively. Bracket counts are computed with multiplicity from powers: $\langle ij \rangle^n$ counts as $n$ brackets, and brackets in both numerator and denominator are included. Source-relative success means the output is strictly simpler than the input in at least one of brackets or terms (without increasing the other). Target-relative success means both the bracket count and the term count are $\leq$ the target's values. These criteria are applied uniformly to both models. For our model, we check whether the final output satisfies them; for CDS, we re-evaluated the pretrained seq2seq models~\cite{CDScode} on their published test sets using beam search with $B{=}20$ (the highest setting considered by CDS), and check whether \emph{any} valid beam hypothesis satisfies the criterion.

Table~\ref{tab:amplitude_results} shows failures under both criteria, broken down by the number of target terms. Under the looser source-relative criterion, our model has a failure rate of only $3.0\times 10^{-4}$ for 4-point (3 failures out of 9,869), $6.8\times 10^{-4}$ for 5-point (7/10,256), and $1.6\times 10^{-3}$ for 6-point (16/9,934). Under the stricter target-relative criterion, these rise modestly to 11, 36, and 63 failures respectively---still below 0.7\% in all cases despite an action space of 29,760 for 6-point, a $660\times$ increase over the dilogarithm case. Meanwhile, the seq2seq model of CDS has significantly higher failure rates, ranging from 1.2--3.9\% under the source-relative criterion to 1.8--4.0\% under the target-relative criterion. Notably, the gap between CDS's source- and target-relative failures is small (e.g., 119 vs.\ 176 for 4-point), indicating that when beam search produces a valid simplification, it typically achieves target-level complexity.

\begin{table}[tb]
\centering
\begin{tabular}{lc cc cc}
\hline\hline
& & \multicolumn{2}{c}{Our model} & \multicolumn{2}{c}{CDS} \\
& $N$ & Source-rel & Target-rel & Source-rel & Target-rel \\
\hline
\multicolumn{6}{l}{\emph{4-point, by target terms}} \\
 0  & 1,338 & 0  & 1  & 0   & 3 \\
 1  & 4,775 & 0  & 0  & 9   & 32 \\
 2  & 2,366 & 1  & 2  & 28  & 53 \\
 3  & 1,390 & 2  & 8  & 82  & 88 \\
Total (4pt) & 9,869  & 3  & 11 & 119 & 176 \\
\hline
\multicolumn{6}{l}{\emph{5-point, by target terms}} \\
 0  & 835   & 0  & 0  & 1   & 1 \\
 1  & 4,629 & 1  & 1  & 29  & 35 \\
 2  & 2,665 & 3  & 5  & 107 & 114 \\
 3  & 2,127 & 3  & 30 & 262 & 264 \\
Total (5pt) & 10,256 & 7  & 36 & 399 & 414 \\
\hline
\multicolumn{6}{l}{\emph{6-point, by target terms}} \\
 0  & 684   & 0  & 0  & 2   & 2 \\
 1  & 4,213 & 5  & 7  & 34  & 35 \\
 2  & 2,733 & 3  & 9  & 88  & 89 \\
 3  & 2,304 & 8  & 47 & 178 & 178 \\
Total (6pt) & 9,934  & 16 & 63 & 302 & 304 \\
\hline\hline
\end{tabular}
\caption{Scattering amplitude simplification failures on the test sets of CDS, excluding pathological samples where the target is not strictly simpler than the source (123 for 4pt, 5 for 5pt, 0 for 6pt) and train/test overlap samples (28 for 4pt, 9 for 5pt, 0 for 6pt), broken down by number of target terms. Bracket counts are power-aware ($\langle ij \rangle^n$ counts as $n$). Both criteria are applied uniformly to both models. Source-relative: failure to produce output strictly simpler in at least one of brackets or terms (without increasing the other). Target-relative: failure to produce output with both bracket count and term count $\leq$ the known target. For CDS, success requires any valid beam hypothesis ($B{=}20$) to satisfy the criterion.}
\label{tab:amplitude_results}
\end{table}

\begin{figure*}[tb]
\centering
\includegraphics[width=0.33\textwidth]{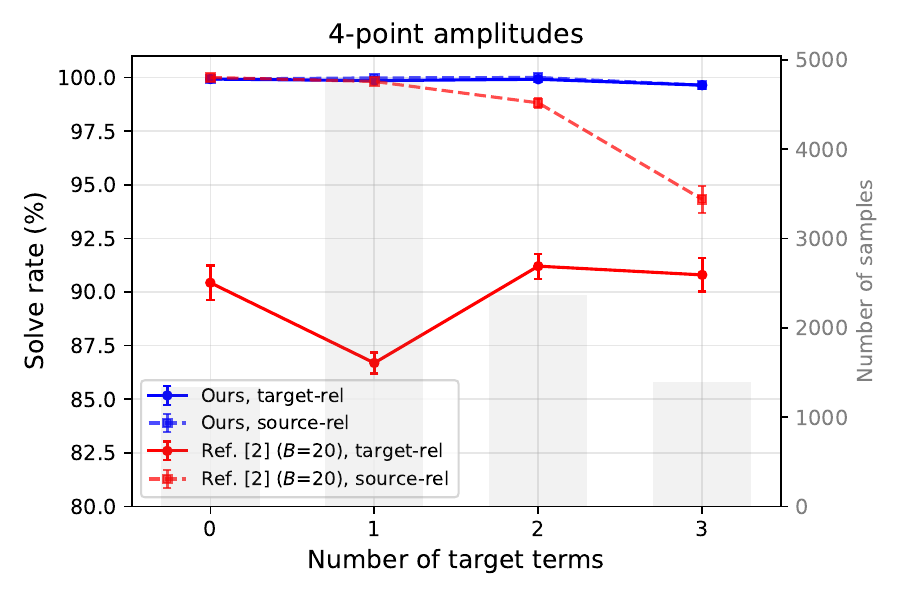}%
\includegraphics[width=0.33\textwidth]{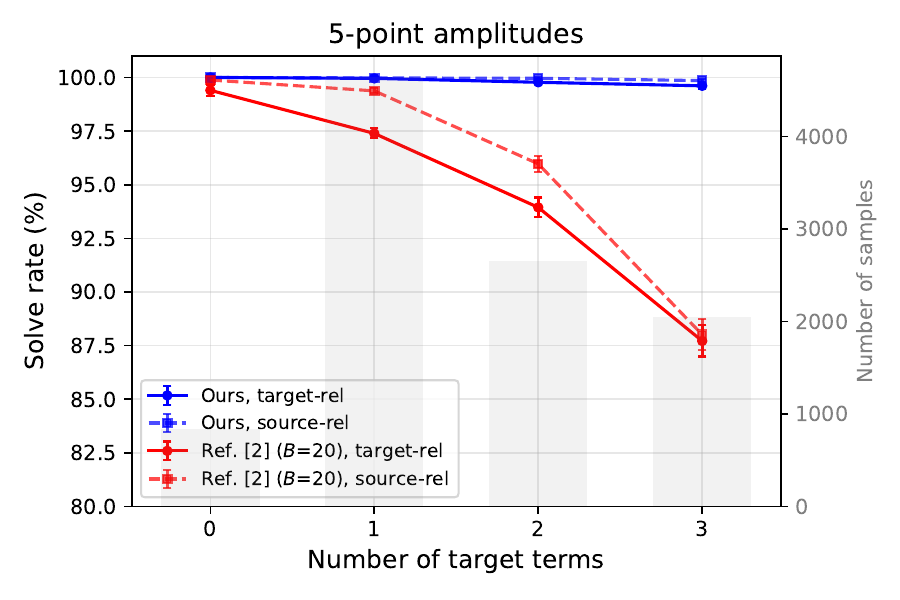}%
\includegraphics[width=0.33\textwidth]{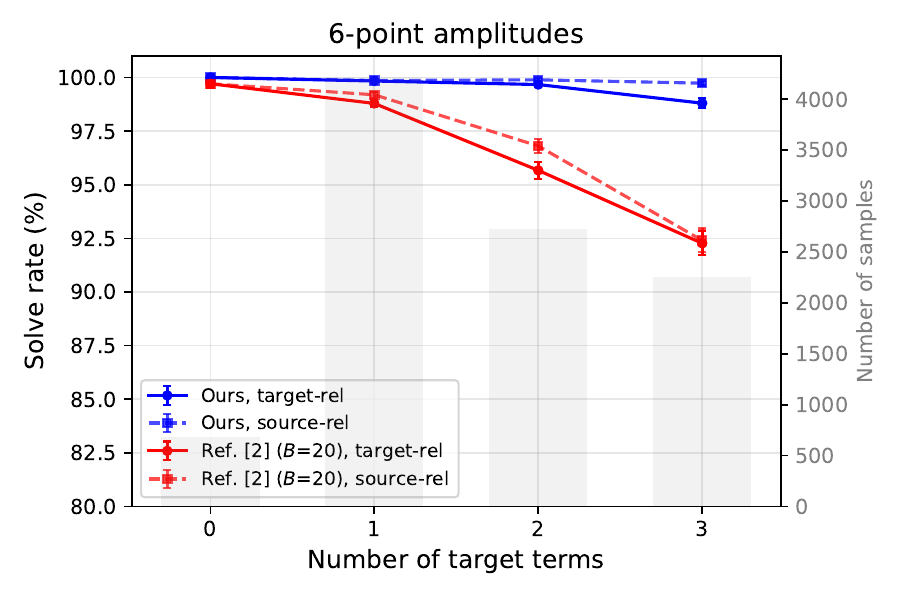}
\caption{Solve rate vs.\ number of target terms for 4-point (left), 5-point (center), and 6-point (right) amplitudes. Performance is shown under source-relative (dashed) and target-relative (solid) criteria for our model (blue) and CDS with $B{=}20$ (red). Gray bars show the number of test samples at each target term count.}
\label{fig:amp_vs_tgt_terms}
\end{figure*}

\begin{figure*}[tb]
\centering
\includegraphics[width=0.33\textwidth]{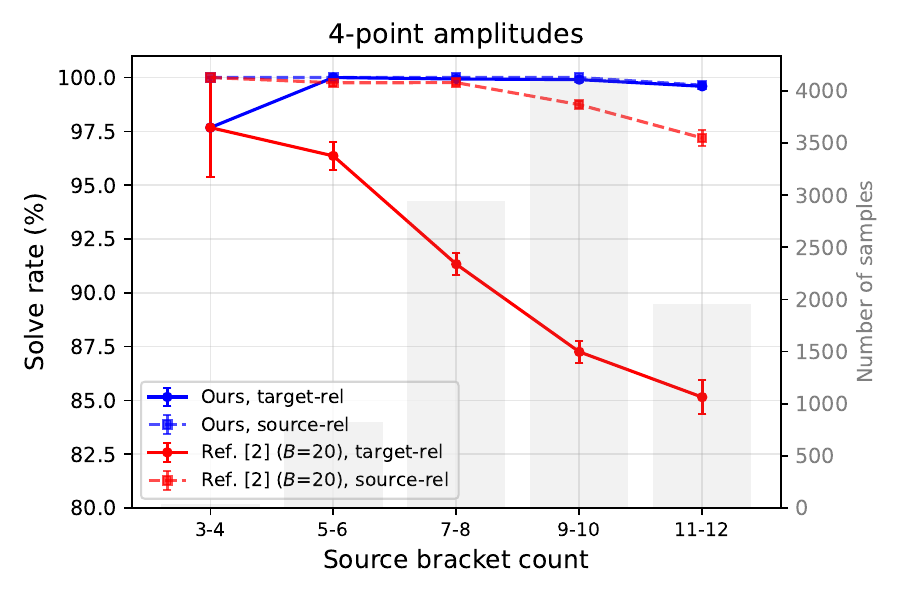}%
\includegraphics[width=0.33\textwidth]{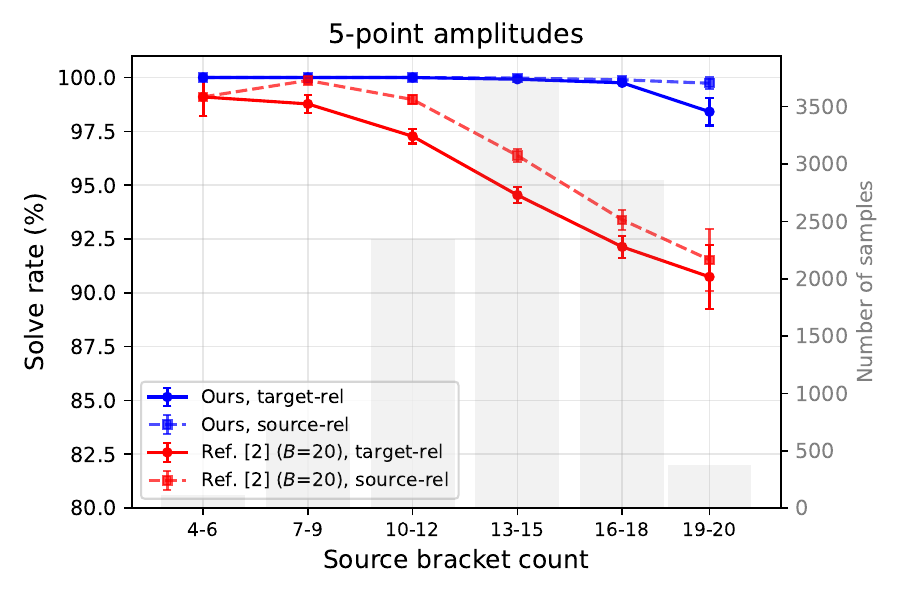}%
\includegraphics[width=0.33\textwidth]{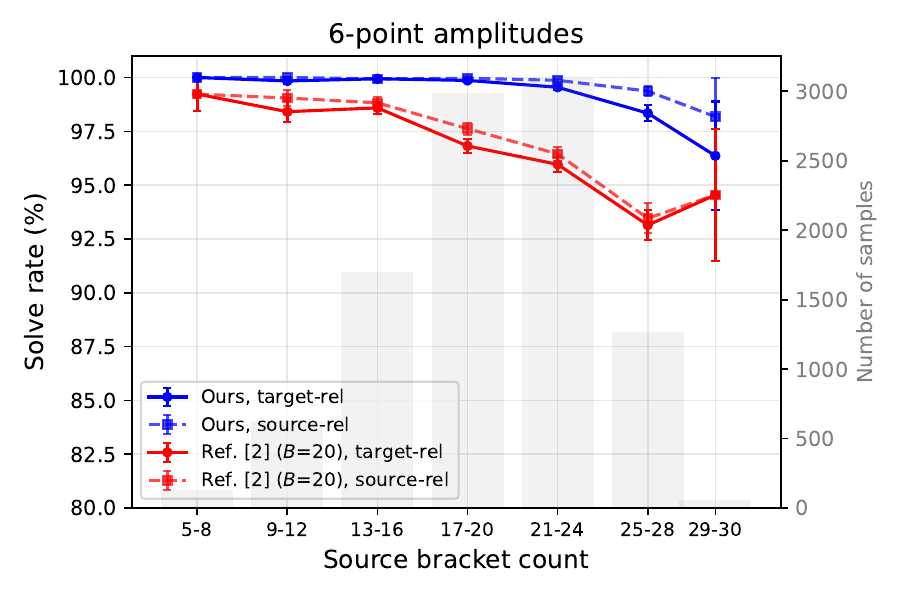}
\caption{Solve rate vs.\ source bracket count for 4-point (left), 5-point (center), and 6-point (right) amplitudes. Our model (blue) maintains near-100\% solve rates across all starting complexities under both source-relative (dashed) and target-relative (solid) criteria. The model of CDS with beam size $B{=}20$ (red) degrades with increasing bracket count. Gray bars show the number of test samples at each bracket count.}
\label{fig:amp_vs_brackets}
\end{figure*}

Table~\ref{tab:amplitude_results}  and  Fig.~\ref{fig:amp_vs_tgt_terms} show that performance of our model is robust across target complexities, and the gap between our model and CDS widens with increasing number of target terms, particularly under the target-relative criterion. Our model's failures are concentrated at higher target term counts, with 47 of the 63 six-point target-relative failures occurring at $n_{\rm tgt}{=}3$.

Performance is also robust across all starting complexities, as shown in Fig.~\ref{fig:amp_vs_brackets}. Our model maintains near-perfect solve rates across all starting bracket counts, with failures concentrated at the highest complexities. The failures are predominantly due to term count exceeding the target rather than bracket count.

Per-sample comparison with CDS reveals that the two models' failure sets are nearly disjoint: our model solves $>$96\% of their target-relative failures (97.2\% for 4pt, 98.1\% for 5pt, 96.4\% for 6pt), and only 5--11 expressions per multiplicity are unsolved by either model. This indicates strong complementarity between the two approaches and suggests a future hybrid approach could be even more successful.

\subsection{Simplifying Tree-Level Gluon Amplitudes in Yang-Mills Theory}
\label{sec:ym5pt}

The results above demonstrate near-perfect simplification on the test sets of CDS, which contain expressions averaging ${\sim}$4, ${\sim}$7, and ${\sim}$10 source terms for 4-, 5-, and 6-point amplitudes respectively. We now turn to a significantly harder challenge: simplifying 5-point tree-level gluon amplitudes in Yang-Mills theory obtained from Feynman diagram calculations.\footnote{We thank Aur\'elien Dersy for providing these amplitudes (private communication).} These expressions (1024 in total) differ only by the choice of assignments of polarization reference spinors to external legs, and range from 8 to 228 numerator terms (mean ${\sim}$90), far exceeding both the CDS test set complexity and the model's input capacity of 25 terms. Each is mathematically equivalent to the Parke-Taylor formula~\eqref{eq:parke_taylor}, which after cancellation reduces to
\begin{equation}
A(1^- 2^- 3^+ 4^+ 5^+) = \frac{\langle 12 \rangle^3}{\langle 15 \rangle \langle 23 \rangle \langle 34 \rangle \langle 45 \rangle}\,,
\label{eq:parke_taylor_5pt}
\end{equation}
a single-term expression with 7 brackets (counting $\langle 12\rangle^3$ with multiplicity). Reaching this compact form from the Feynman diagram output requires a long sequence of identity applications through a vast combinatorial search space. We select a representative subset of 103 forms (every 10th form, indexed 0 through 1020) and attempt to simplify each one to the Parke-Taylor target.

Our simplification pipeline consists of two phases: a greedy contrastive-grouping phase that reduces large expressions to manageable size, followed by a beam search phase that completes the simplification.

\subsubsection{Phase 1: Greedy Contrastive Grouping}

Since input expressions have up to ${\sim}$200 terms but the model accepts at most 25, we follow CDS and use a \emph{contrastive grouping} strategy to decompose the problem. A pre-trained contrastive encoder~\cite{Cheung:2024tbb, CDScode} computes an embedding for each numerator term, and pairwise cosine similarities identify groups of terms likely to simplify together. Each group forms a sub-expression small enough for the model.

At each step, the procedure:
\begin{enumerate}
\item Computes the pairwise cosine similarity matrix over all current numerator terms (with numerical coefficients removed for robustness).
\item For each reference term, selects the most similar neighbors above a threshold, forming groups of up to 25 terms.
\item For each group, factors out common spinor brackets shared by all terms, reducing the sub-expression complexity before model evaluation.
\item Applies a single model action to the reduced sub-expression, accepts the result if it reduces the term count, and reassembles with the remaining terms.
\item After a full pass through all reference terms, re-cancels the expression and begins a new pass with updated similarities.
\end{enumerate}
The similarity threshold starts at 0.6 and relaxes across passes (through 0.5 and 0.4), progressively allowing less similar terms to be grouped. Up to 100 passes are performed per form, with numerical validation after each pass to detect errors.

This greedy phase is fast and is effective at reducing large expressions: from a mean of 90 terms, it achieves a mean output of ${\sim}$32 terms. Of the 103 forms, 17 are solved directly to the 1-term Parke-Taylor formula in this phase alone, including forms with up to 146 input terms.

\subsubsection{Phase 2: Beam Search}

Forms not solved by the greedy phase enter a beam search over sequences of identity applications. The beam maintains the top $K{=}20$ candidate expressions, ranked by $(n_{\rm terms}, n_{\rm brackets})$. At each step, every beam entry is expanded using contrastive grouping: the contrastive encoder computes pairwise cosine similarities among all numerator terms, and for each reference term and each similarity threshold (0.6, 0.5, 0.4), the most similar neighbors are grouped into a sub-expression of at most 25 terms. All unique groupings across reference terms and thresholds are enumerated, and for each grouping, the model's highest-scoring valid action is applied to the sub-expression and the result is recombined with the remaining terms. This produces many candidate children per beam entry at each step (one per unique grouping), all of which compete for the $K{=}20$ beam slots.
Key design choices:
\begin{itemize}
\item \textbf{Deduplication.} A global set of visited expression hashes (using deterministic MD5 hashing) prevents the beam from cycling back to previously explored states.
\item \textbf{Parent carryover.} All current beam entries are carried forward as candidates at each step, ensuring the best-known expression is never lost.
\item \textbf{Progressive exploration.} Each beam entry tracks how many actions have been tried on it; when a parent survives to the next step, the worker skips previously attempted actions and explores new ones.
\item \textbf{Patience.} The search terminates if no improvement (reduction in best term or bracket count) occurs for 20 consecutive steps.
\end{itemize}
The beam search runs for up to 200--300 steps per form, with each step parallelized across HTCondor workers for the encoding and simplification sub-tasks.

\subsubsection{Results}

All 103 forms are successfully simplified to the 1-term Parke-Taylor formula, a 100\% solve rate. Of these, 17 are solved in the greedy phase alone and the remaining 86 require beam search. The hardest form (form990, starting at 178 terms) required 98 beam search steps to reach the target.

\begin{figure}[t]
\centering
\includegraphics[width=\columnwidth]{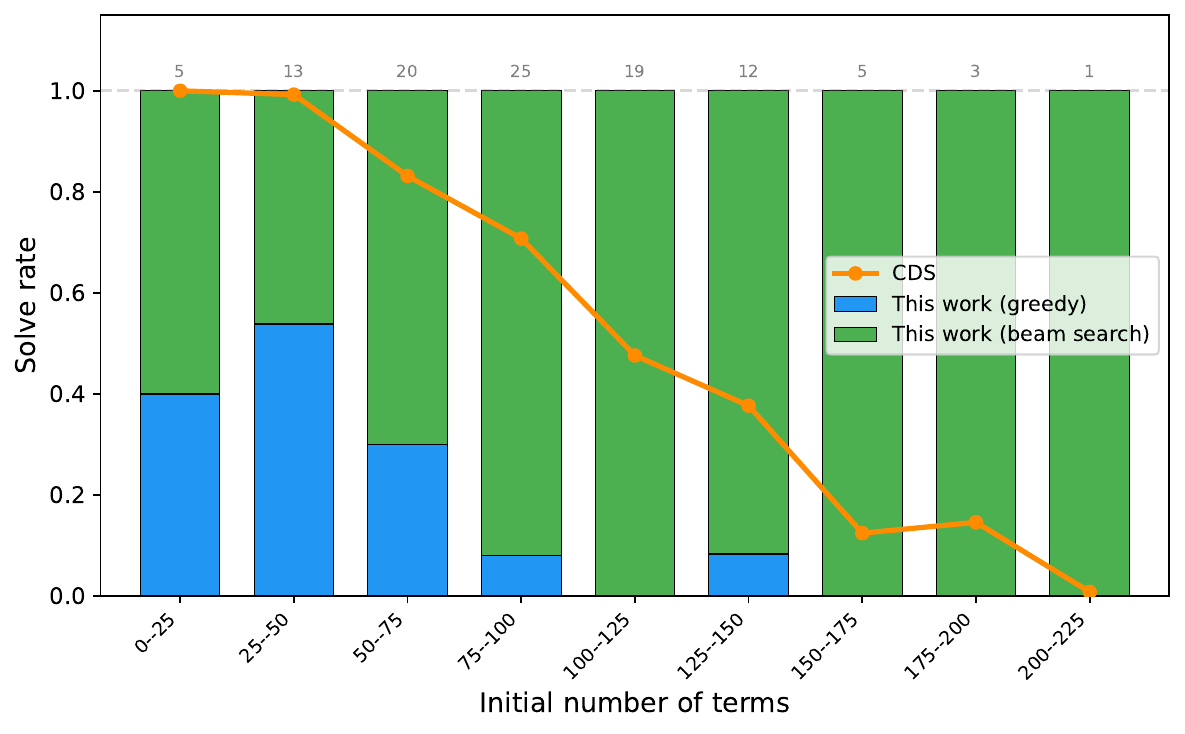}
\caption{Solve rate for 5-point Yang-Mills partial amplitudes as a function of the initial number of numerator terms, binned in groups of 25. Stacked bars show our results: blue indicates forms solved by the greedy contrastive-grouping phase alone, green indicates forms additionally requiring beam search. The orange curve shows the CDS sequential simplification solve rate, digitized from Fig.~9 (right panel) of Ref.~\cite{Cheung:2024tbb} and averaged within each bin. Numbers above each bar indicate the number of forms in that bin. Our method achieves a 100\% solve rate across all bins (103/103 forms), while CDS performance degrades with expression complexity.}
\label{fig:ym5pt_solve_rate}
\end{figure}

Fig.~\ref{fig:ym5pt_solve_rate} compares our solve rate against CDS as a function of the initial number of terms. Our method achieves 100\% across all complexity bins, while the CDS sequential simplification (digitized from Fig.~9 of Ref.~\cite{Cheung:2024tbb}) declines sharply with expression size, dropping below 50\% for expressions with more than 100 terms.

These results demonstrate that our trained model, combined with contrastive grouping to handle expressions beyond its input capacity and beam search to navigate the combinatorial space of identity sequences, can simplify realistic Feynman-diagram-level amplitudes to their known compact forms.

\section{Discussion and Conclusions}
\label{sec:discussion}

We have presented a self-supervised learning approach for symbolic simplification in which oracle trajectories are generated by scrambling simple expressions and reversing the operations. Combined with a multi-label loss to handle action equivalence and inference-time techniques (anti-cycle detection, backtracking, reject-term-increase), this achieves near-perfect performance on two challenging problems: 99.9\% on dilogarithm reduction (vs.\ 92\% for DSZ) and 99.4--99.9\% on spinor-helicity amplitude simplification (vs.\ 96.0--98.2\% for CDS under the target-relative criterion). The approach scales well from 45 actions (dilogarithms) to 29,760 actions (6-point amplitudes), and generalizes beyond its training distribution: the dilogarithm model achieves 99.9\% on expressions scrambled up to 10 times despite training on at most 7 scrambles. Furthermore, by combining the trained MDP with contrastive grouping and beam search, we achieve a 100\% solve rate on tree-level 5-point gluon amplitudes with up to 228 terms---far beyond the model's training distribution---demonstrating that the learned per-step policy can generalize to simplify realistic Feynman-diagram-level expressions.

We believe the key advantage of our approach over end-to-end regression is the decomposition of a difficult global mapping into a sequence of learnable local steps, with oracle trajectories providing the supervision signal for each step. This is also what enables generalization beyond the training distribution: the model learns \emph{per-transition} decisions that compose regardless of total expression complexity. Compared to RL, the combination of deterministic dynamics, known goal states, and reversible scrambling operations enables unlimited generation of expert demonstrations, bypassing the sparse-reward and sample-efficiency challenges that make RL difficult in this setting.

The success of this work opens up many potential future directions. Within the polylogarithm domain, extending to higher transcendental weight functions through the use of scrambled symbols as in \cite{Dersy:2022ltg} should be natural and straightforward, potentially building on tools like PolyLogTools~\cite{DuhrDulat2019}. Extending to the problem of integrating symbols back into functions is an additional challenge that would be very interesting to study and may connect back to the more general problem developing ML methods for symbolic integration \cite{Lample2019}. For amplitudes, the extension beyond spinor-helicity would have much practical benefit, in particular to loop integrals and the reduction to master integrals with integration-by-parts identities~\cite{Chetyrkin:1981qh,Laporta:2001dd} (see \cite{vonHippel:2025okr,Song:2025ibp,Zeng:2025ibp} for other ML methods recently applied to IBP reduction). Recent work applying transformers directly to amplitude computation~\cite{Cai2024,Cai2025} suggests further synergies between ML and the amplitudes program. More broadly, the scramble-and-reverse paradigm should apply to any symbolic domain with reversible rewrite rules and known simple forms. Finally, the two examples we studied here turned out to be almost completely solvable with just the self-supervised oracle trajectories. It would be interesting to explore if there are harder settings where the self-supervision is insufficient. Then perhaps our oracle trajectory framework can be combined with reinforcement learning,  perhaps using the self-supervised policy as a warm start for further RL fine-tuning. 

\section*{Acknowledgements}
I am grateful to Aurelien Dersy for helpful conversations and feedback on the draft, and to Marat Freytsis and Siddharth Mishra-Sharma for making this work possible by facilitating the initial access to Claude Max.  This work was done in full collaboration with Claude Code Opus 4.5-4.6. Claude did all of the hands on work under my supervision -- writing the code; training, evaluating and optimizing the ML algorithms; analyzing the results; making plots and tables; and contributing significantly to the writing of the paper. It also made meaningful contributions to brainstorming and problem solving. This research was supported by DOE grant DOE-SC0010008 and was supported in part by grant NSF PHY-2309135 to the Kavli Institute for Theoretical Physics (KITP). This work was performed in part at the Aspen Center for Physics, which is supported by National Science Foundation grant PHY-2210452.

\section*{Code}
Our code is available at \url{https://github.com/davidshih17/RL_dilogs_claude} and \url{https://github.com/davidshih17/RL_amplitudes_claude}.

\appendix

\section{Dilogarithm Failure Cases}
\label{app:dilog_failures}

The 6 target-relative failures of our model on the clean dilogarithm test set are listed below. For each, we show: the source expression (input), our model's best output (minimum term count achieved during the rollout), and the known target. Failures marked with $^*$ are also source-relative failures (no term reduction at all).

\medskip
\noindent\textbf{Failure 1} ($n_s{=}3$, depth 9, $11 \to 5$ terms):
{\small
\begin{align}
\text{src} &= -7\,\text{Li}_2\!\big(\!\!-\tfrac{1}{x}\big) - 7\,\text{Li}_2\!\big(\tfrac{x^2+2x+2}{x+2}\big) \notag \\
&\quad + 4\,\text{Li}_2\!\big(\tfrac{3x-4}{2x^2+x-2}\big) - 3\,\text{Li}_2\!\big(\tfrac{2x^2-1}{x^2-x+1}\big) \notag \\
&\quad - 4\,\text{Li}_2\!\big(\tfrac{2x^2+x-2}{2x^2-2x+2}\big) + 7\,\text{Li}_2(x{+}1) \notag \\
&\quad - 3\,\text{Li}_2\!\big(\!\!-\tfrac{x^2}{2}{-}\tfrac{x}{2}{+}1\big) - 3\,\text{Li}_2\!\big(\tfrac{x^2}{2}{+}\tfrac{x}{2}{-}1\big) \notag \\
&\quad - 3\,\text{Li}_2(x^2{+}2x{-}1) \notag \\
&\quad - \tfrac{3}{2}\,\text{Li}_2\!\big(\tfrac{4}{x^4+2x^3-3x^2-4x+4}\big) \notag \\
&\quad + \tfrac{3}{2}\,\text{Li}_2(x^4{+}4x^3{+}2x^2{-}4x{+}1) \notag \\
\text{ours} &= -7\,\text{Li}_2\!\big(\!\!-\tfrac{1}{x}\big) - 7\,\text{Li}_2\!\big(\tfrac{x^2+2x+2}{x+2}\big) \notag \\
&\quad - 3\,\text{Li}_2\!\big(\tfrac{2x^2-1}{x^2-x+1}\big) + 3\,\text{Li}_2\!\big(\tfrac{1}{x^2+2x}\big) \notag \\
&\quad - 7\,\text{Li}_2\!\big(\tfrac{1}{x+1}\big) \notag \\
\text{tgt} &= -7\,\text{Li}_2\!\big(\tfrac{x^2+2x+2}{x+2}\big) \notag \\
&\quad - 3\,\text{Li}_2\!\big(\tfrac{-x^2+x-1}{x^2+x-2}\big) - 3\,\text{Li}_2(x^2{+}2x) \notag
\end{align}
}

\noindent\textbf{Failure 2} ($n_s{=}2$, depth 7, $11 \to 7$ terms):
{\small
\begin{align}
\text{src} &= -8\,\text{Li}_2\!\big(\tfrac{1}{2x-1}\big) - 8\,\text{Li}_2\!\big(\tfrac{x^2+2}{2x+2}\big) \notag \\
&\quad - 7\,\text{Li}_2\!\big(\tfrac{2}{2x+1}\big) - 7\,\text{Li}_2\!\big(\tfrac{2x-1}{2x+1}\big) \notag \\
&\quad - 6\,\text{Li}_2\!\big(1{-}\tfrac{x}{2}\big) - 4\,\text{Li}_2(2x{-}1) \notag \\
&\quad - 2\,\text{Li}_2(16x^4{-}32x^3{+}24x^2{-}8x{+}1) \notag \\
&\quad + 2\,\text{Li}_2(4x^2{-}4x{+}1) + 4\,\text{Li}_2(2x) \notag \\
&\quad + 4\,\text{Li}_2(-4x^2{+}4x{-}1) + 8\,\text{Li}_2\!\big(\tfrac{2x}{2x-1}\big) \notag \\
\text{ours} &= -8\,\text{Li}_2\!\big(\tfrac{1}{2x-1}\big) - 8\,\text{Li}_2\!\big(\tfrac{x^2+2}{2x+2}\big) \notag \\
&\quad - 6\,\text{Li}_2\!\big(1{-}\tfrac{x}{2}\big) - 4\,\text{Li}_2(2x{-}1) \notag \\
&\quad - 2\,\text{Li}_2(4x^2{-}4x{+}1) - 4\,\text{Li}_2(1{-}2x) \notag \\
&\quad - 8\,\text{Li}_2\!\big(\tfrac{2x-1}{2x}\big) \notag \\
\text{tgt} &= 6\,\text{Li}_2\!\big(\tfrac{x}{2}\big) - 8\,\text{Li}_2\!\big(\tfrac{x^2+2}{2x+2}\big) \notag
\end{align}
}

\noindent\textbf{Failure 3}$^*$ ($n_s{=}2$, depth 6, $4 \to 4$ terms, no reduction):
{\small
\begin{align}
\text{src} &= 5\,\text{Li}_2\!\big(\tfrac{-x^2+3x+3}{x^2-x-1}\big) + 6\,\text{Li}_2\!\big(\tfrac{1-x}{x}\big) \notag \\
&\quad - \tfrac{5}{2}\,\text{Li}_2\!\big(\tfrac{-4x^3+4x^2+16x+8}{x^4-2x^3-x^2+2x+1}\big) \notag \\
&\quad - \tfrac{5}{4}\,\text{Li}_2\!\bigg(\tfrac{x^8{-}4x^7{-}14x^6{+}40x^5{+}107x^4{-}72x^3{-}318x^2{-}252x{-}63}{x^8{-}4x^7{+}2x^6{+}8x^5{-}5x^4{-}8x^3{+}2x^2{+}4x{+}1}\bigg) \notag \\
\text{ours} &= -5\,\text{Li}_2\!\big(\tfrac{x^2-3x-3}{x^2-x-1}\big) + 6\,\text{Li}_2\!\big(\tfrac{1-x}{x}\big) \notag \\
&\quad - \tfrac{5}{2}\,\text{Li}_2\!\big(\tfrac{-x^4+6x^3-3x^2-18x-9}{x^4-2x^3-x^2+2x+1}\big) \notag \\
&\quad - \tfrac{5}{2}\,\text{Li}_2\!\big(\tfrac{-2x^4+6x^2+12x+6}{x^4-2x^3-x^2+2x+1}\big) \notag \\
&\quad + \tfrac{5}{4}\,\text{Li}_2\!\bigg(\tfrac{x^8{-}12x^7{+}42x^6{-}189x^4{+}378x^2{+}324x{+}81}{x^8{-}4x^7{+}2x^6{+}8x^5{-}5x^4{-}8x^3{+}2x^2{+}4x{+}1}\bigg) \notag \\
&\quad + \tfrac{5}{4}\,\text{Li}_2\!\bigg(\tfrac{4x^8{-}24x^6{-}48x^5{+}12x^4{+}144x^3{+}216x^2{+}144x{+}36}{x^8{-}4x^7{+}2x^6{+}8x^5{-}5x^4{-}8x^3{+}2x^2{+}4x{+}1}\bigg) \notag \\
&\quad + \tfrac{5}{4}\,\text{Li}_2\!\bigg(\tfrac{x^8{+}4x^7{-}6x^6{-}48x^5{-}45x^4{+}112x^3{+}266x^2{+}196x{+}49}{x^8{-}4x^7{+}2x^6{+}8x^5{-}5x^4{-}8x^3{+}2x^2{+}4x{+}1}\bigg) \notag \\
&\quad - \tfrac{5}{4}\,\text{Li}_2\!\bigg(\tfrac{x^8{-}4x^7{-}14x^6{+}40x^5{+}107x^4{-}72x^3{-}318x^2{-}252x{-}63}{x^8{-}4x^7{+}2x^6{+}8x^5{-}5x^4{-}8x^3{+}2x^2{+}4x{+}1}\bigg) \notag \\
\text{tgt} &= -6\,\text{Li}_2\!\big(\tfrac{-x}{x-1}\big) - 5\,\text{Li}_2\!\big(\tfrac{x^2-x-1}{2x+2}\big) \notag
\end{align}
}

\noindent\textbf{Failure 4} ($n_s{=}1$, depth 10, $10 \to 7$ terms):
{\small
\begin{align}
\text{src} &= -7\,\text{Li}_2\!\big(\!\!-\tfrac{1}{x^2+x}\big) - 7\,\text{Li}_2(-x^2{-}x) \notag \\
&\quad - 3\,\text{Li}_2\!\big(\!\!-\tfrac{x}{2}{-}1\big) - 3\,\text{Li}_2(-2x{-}2) \notag \\
&\quad - 3\,\text{Li}_2\!\big(\tfrac{x}{2}{+}2\big) - 3\,\text{Li}_2(2x{+}2) \notag \\
&\quad + 3\,\text{Li}_2\!\big(\tfrac{1}{x^4+1}\big) + 6\,\text{Li}_2\!\big(\tfrac{x^2-1}{x^2}\big) \notag \\
&\quad - \tfrac{3}{2}\,\text{Li}_2(x^8) + \tfrac{3}{2}\,\text{Li}_2(4x^2{+}8x{+}4) \notag \\
\text{ours} &= 7\,\text{Li}_2\!\big(\tfrac{1}{x^2+x}\big) \notag \\
&\quad + \tfrac{7}{2}\,\text{Li}_2\!\big(\!\!-\tfrac{1}{x^4+2x^3+x^2}\big) \notag \\
&\quad - \tfrac{7}{4}\,\text{Li}_2\!\big(\tfrac{1}{x^8+4x^7+6x^6+4x^5+x^4}\big) \notag \\
&\quad - 7\,\text{Li}_2(-x^2{-}x) - 3\,\text{Li}_2\!\big(\!\!-\tfrac{x}{2}{-}1\big) \notag \\
&\quad - 3\,\text{Li}_2\!\big(\tfrac{x}{2}{+}2\big) + 3\,\text{Li}_2\!\big(\tfrac{1}{x^4+1}\big) \notag \\
&\quad + 6\,\text{Li}_2\!\big(\tfrac{x^2-1}{x^2}\big) + \tfrac{3}{2}\,\text{Li}_2(x^{-8}) \notag \\
\text{tgt} &= 6\,\text{Li}_2(x^2{+}1) \notag
\end{align}
}

\noindent\textbf{Failure 5}$^*$ ($n_s{=}1$, depth 7, $3 \to 3$ terms, no reduction):
{\small
\begin{align}
\text{src} &= -4\,\text{Li}_2\!\big(\tfrac{-4x^4-8x^3+4x-1}{3x^4+8x^3-4x^2-4x-3}\big) \notag \\
&\quad - 2\,\text{Li}_2\!\bigg(\tfrac{9x^8{+}48x^7{+}40x^6{-}88x^5{-}66x^4{-}16x^3{+}40x^2{+}24x{+}9}{16x^8{+}64x^7{+}64x^6{-}32x^5{-}56x^4{+}16x^3{+}16x^2{-}8x{+}1}\bigg) \notag \\
&\quad + 8\,\text{Li}_2\!\big(\tfrac{-x^2-2}{x^2+2x-3}\big) \notag \\
\text{ours} &= 4\,\text{Li}_2\!\big(\tfrac{-3x^4-8x^3+4x^2+4x+3}{4x^4+8x^3-4x+1}\big) \notag \\
&\quad - 2\,\text{Li}_2\!\bigg(\tfrac{16x^8{+}64x^7{+}64x^6{-}32x^5{-}56x^4{+}16x^3{+}16x^2{-}8x{+}1}{7x^8{+}16x^7{+}24x^6{+}56x^5{+}10x^4{+}32x^3{-}24x^2{-}32x{-}8}\bigg) \notag \\
&\quad - 8\,\text{Li}_2\!\big(\tfrac{2x^2+2x-1}{x^2+2x-3}\big) \notag \\
\text{tgt} &= -8\,\text{Li}_2\!\big(\tfrac{-2x^2-2x+1}{x^2+2}\big) \notag
\end{align}
}

\noindent\textbf{Failure 6} ($n_s{=}0$, depth 9, $5 \to 2$ terms):
{\small
\begin{align}
\text{src} &= -3\,\text{Li}_2\!\big(\tfrac{2x^2-x-2}{x^2+x-2}\big) + 3\,\text{Li}_2\!\big(\tfrac{x^2-2x}{2x^2-x-2}\big) \notag \\
&\quad - \tfrac{3}{2}\,\text{Li}_2\!\big(\tfrac{3x^4-6x^3-4x^2+8x}{4x^4-4x^3-7x^2+4x+4}\big) \notag \\
&\quad - \tfrac{3}{2}\,\text{Li}_2\!\big(\tfrac{x^4+2x^3-3x^2-4x+4}{5x^4-2x^3-10x^2+8}\big) \notag \\
&\quad - \tfrac{3}{4}\,\text{Li}_2\!\bigg(\tfrac{x^8{+}4x^7{-}2x^6{-}20x^5{+}x^4{+}40x^3{-}8x^2{-}32x{+}16}{16x^8{-}32x^7{-}40x^6{+}88x^5{+}49x^4{-}88x^3{-}40x^2{+}32x{+}16}\bigg) \notag \\
\text{ours} &= -3\,\text{Li}_2\!\big(\tfrac{2x^2-x-2}{x^2+x-2}\big) - 3\,\text{Li}_2\!\big(\tfrac{-x^2+2x}{x^2+x-2}\big) \notag \\
\text{tgt} &= 0 \notag
\end{align}
}

\bibliographystyle{apsrev4-1}
\bibliography{refs}

\end{document}